\documentclass[conference]{IEEEtran}
\makeatletter
\def\endthebibliography{%
  \def\@noitemerr{\@latex@warning{Empty `thebibliography' environment}}%
  \endlist
}
\makeatother
\IEEEoverridecommandlockouts
\usepackage{cite}
\usepackage{amsmath,amssymb,amsfonts, bm}
\usepackage{algorithmic}
\usepackage[ruled, linesnumbered]{algorithm2e}
\usepackage{graphicx}
\usepackage{textcomp}
\usepackage{xcolor}
\usepackage{bbm}
\usepackage{dsfont}
\usepackage{hyperref}
\usepackage{url}
\usepackage{fancyhdr}

\def\BibTeX{{\rm B\kern-.05em{\sc i\kern-.025em b}\kern-.08em
    T\kern-.1667em\lower.7ex\hbox{E}\kern-.125emX}}

\makeatletter 
\newcommand{\linebreakand}{%
  \end{@IEEEauthorhalign}
  \hfill\mbox{}\par
  \mbox{}\hfill\begin{@IEEEauthorhalign}
}
\makeatother 

\begin{document}

\title{Differentiable Bootstrap Particle Filters \\ for Regime-Switching Models}

\author{\IEEEauthorblockN{Wenhan Li}
\IEEEauthorblockA{\textit{Department of Computer Science} \\
\textit{University of Surrey}\\
Guildford, UK \\
wenhan.li@surrey.ac.uk}
\and
\IEEEauthorblockN{Xiongjie Chen}
\IEEEauthorblockA{\textit{Department of Computer Science} \\
\textit{University of Surrey}\\
Guildford, UK \\
xiongjie.chen@surrey.ac.uk}
\and
\IEEEauthorblockN{Wenwu Wang}
\IEEEauthorblockA{\textit{Centre for Vision, Speech and Signal Processing} \\
\textit{University of Surrey}\\
Guildford, UK \\
w.wang@surrey.ac.uk}
\and

\linebreakand

\IEEEauthorblockN{V{\'i}ctor Elvira}
\IEEEauthorblockA{\textit{School of Mathematics} \\
\textit{University of Edinburgh}\\
Edinburgh, UK \\
victor.elvira@ed.ac.uk}
\and
\IEEEauthorblockN{Yunpeng Li}
\IEEEauthorblockA{\textit{Department of Computer Science} \\
\textit{University of Surrey}\\
Guildford, UK \\
yunpeng.li@surrey.ac.uk}
}


\maketitle
\thispagestyle{fancy}
\fancyhead{}
\lhead{}
\lfoot{978-1-6654-5245-8/23/\textdollar31.00 ©2023 IEEE}
\cfoot{}
\rfoot{}

\begin{abstract}
Differentiable particle filters are an emerging class of particle filtering methods that use neural networks to construct and learn parametric state-space models. In real-world applications, both the state dynamics and measurements can switch between a set of candidate models. For instance, in target tracking, vehicles can idle, move through traffic, or cruise on motorways, and measurements are collected in different geographical or weather conditions. This paper proposes a new differentiable particle filter for regime-switching state-space models. The method can learn a set of unknown candidate dynamic and measurement models and track the state posteriors. We evaluate the performance of the novel algorithm in relevant models, showing its great performance compared to other competitive algorithms.  

\end{abstract}


\section{Introduction}

Inferring unknown quantities based on sequential observations is an important task in many real-world data analysis problems.  One common example is Bayesian filtering, which aims to sequentially estimate posterior distributions of hidden states given observations in a state-space model~\cite{doucet2009tutorial}. Sequential Monte Carlo methods~\cite{djuric2003particle,doucet2001sequential}, a.k.a. particle filters (PFs), are a class of Monte Carlo algorithms where the posteriors are recursively updated and approximated by a set of particles, i.e. weighted Monte Carlo samples. Since the seminal work on the bootstrap particle filter (BPF)~\cite{gordon1993novel}, many variants of particle filters have been proposed, such as the auxiliary particle filter (APF)~\cite{pitt1999filtering,elvira2018search,elvira2019elucidating,branchini2021optimized}, the Gaussian sum particle filter (GSPF)~\cite{kotecha2003gaussian,kotecha2001gaussian}, and the Rao-Blackwellised particle filter (RBPF)~\cite{doucet2000rao,de2002rao}. They are designed for non-linear non-Gaussian filtering tasks where the posteriors are analytically intractable and have been widely used in various real-world applications including geoscience~\cite{van2019particle}, robotics~\cite{gunatilake2022novel}, control systems~\cite{pozna2022hybrid}, and machine learning~\cite{dupty2021pf}.

Particle filters require the knowledge of state evolution (described by a dynamic model) and the link between the hidden state and an observation (via a measurement model). It is often non-trivial to specify these models in real-world applications where complex dynamic patterns and high-dimensional observations exist~\cite{kantas2015particle}. An added layer of complexity is that both the state dynamics and observations can switch between a set of candidate models~\cite{mcginnity2000multiple,liang2009multiple,liu2011instantaneous,urteaga2016sequential,martino2017cooperative,wang2022indoor,el2021particle}. For example, a manoeuvring vehicle can exhibit a mixture of dynamic patterns ranging from moving through city traffic to cruising on motorways. Camera observations in autonomous vehicles are affected by light and weather conditions. This poses an interesting question on how to construct state-space models and perform particle filtering that account for a mixture of switching sub-models.

One class of solutions is to employ a bank of particle filters, one for each candidate model, before fusing the results of each filter~\cite{liu2011instantaneous,urteaga2016sequential,martino2017cooperative,wang2022indoor}. They can incur high computational complexity when the number of candidate models is high. Another direction is to construct regime-switching particle filters~\cite{el2021particle} that augment the state space with the regime index while allowing for a flexible regime index proposal distribution\footnote{Note that we use the terms ``regime'', ``pattern'', ``candidate model'', and ``sub-model'' interchangeably throughout the paper.}. Both classes of methods commonly assume that candidate models either are pre-defined~\cite{liu2011instantaneous,urteaga2016sequential,martino2017cooperative,wang2022indoor,el2021particle} or follow specific model structures so that model parameters can be estimated analytically~\cite{fearnhead2004particle, caron2007bayesian}. This restricts their applicability and effectiveness in real-world filtering tasks.


Differentiable particle filters (DPFs) are a family of recently emerging particle filtering approaches characterised by building and learning components of particle filters with neural networks through automatic differentiation~\cite{karkus2018particle, jonschkowski2018differentiable,ma2020particle,kloss2021train,chen2021differentiable,corenflos2021differentiable}. Several variants~\cite{karkus2018particle, jonschkowski2018differentiable} adopt Gaussian dynamic models due to the simplicity of their differentiable implementations via the reparameterisation trick~\cite{kingma2013auto}. Normalising flows~\cite{rezende2015variational} have been adopted to construct more complicated dynamic models~\cite{chen2021differentiable}. For measurement models, the conditional likelihood of an observation can be obtained as a direct neural network output~\cite{karkus2018particle, jonschkowski2018differentiable}, feature similarity~\cite{wen2021end}, or derived using a conditional normalising flow~\cite{chen2022conditional}. To the best of our knowledge, existing differentiable particle filters have not considered dynamic and measurement models with switching regimes. While generative models such as normalising flows are expressive in theory, it is unclear whether they are effective in practice when coupled with differentiable particle filters in filtering tasks with a set of candidate models. 


In this paper, we propose a new differentiable particle filter algorithm able to learn the models that govern the state dynamics and the observations in regime-switching state-space models. 
The resulted \textit{regime-switching differentiable bootstrap particle filter} combines the best of both worlds -- it inherits the desired properties of regime-switching particle filters including the flexibility to switch between candidate models without running separate filters, with the added benefit to learn unknown candidate models via the optimisation of neural networks. We demonstrate its effectiveness in non-linear filtering simulations with switching regimes. 


The rest of the paper is organised as follows.  Section~\ref{sec:problem} formulates the problem. Related work including regime switching particle filters and differentiable particle filters is introduced in Section~\ref{sec:related}. We present the regime-switching differentiable bootstrap particle filters in Section~\ref{sec:method}.  Section~\ref{sec:experiment} provides simulation results and we conclude the paper in Section~\ref{sec:conclusion}.

\section{Problem formulation}
\label{sec:problem}

\begin{figure}[htbp]
    \centering
    \centerline{\includegraphics[width=8cm]{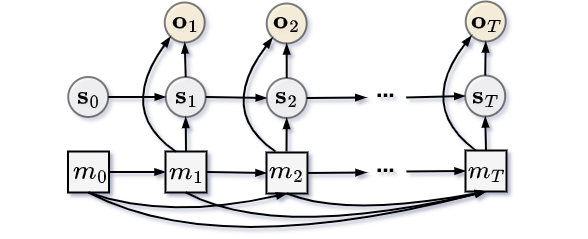}}
    %
    \caption{Diagram of a regime switching state-space model with a period of $T$.}
    \label{fig:diagram}
\end{figure}

We consider nonlinear filtering tasks with switching dynamic and measurement models defined as follows~\cite{el2021particle}: 
\begin{align}
	m_0 &\sim \pi(m_0)\,, \label{eq:model_prior}\\
	m_t &\sim \phi(m_{t}|m_{1:t-1})\,, \\
	\mathbf{s}_0 &\sim \mu(\mathbf{s}_0)\,, \\
	\mathbf{s}_t &\sim f_{\bm{\theta}_{m_t}}(\mathbf{s}_t|\mathbf{s}_{t-1})\,, \\
	\mathbf{o}_t &\sim g_{\bm{\theta}_{m_t}}(\mathbf{o}_t|\mathbf{s}_t)\,, \label{eq:measurement_model}
\end{align}
where $t\in\mathbb{N}^+$ represents the time index, $\bm{\theta}_{m_t}\in\bm{\Theta}$ is the parameter set of the $m_t$-th regime of the dynamic system. The regime indices $\{m_t\}_{t \geq 1}$ take values from a discrete space $\mathcal{M}:=\{1, \cdots, N_m\}$, and are distributed according to a categorical distribution $m_t\sim \mathcal{C} (p(m_t=1|m_{1:t-1}), \cdots, p(m_t=N_m|m_{1:t-1}))$.  In a time-varying latent Markov process $\{\mathbf{s}_t\}_{t \geq 1}$, the $d_{\mathbf{s}}$-dimensional hidden state of interest $\mathbf{s}_t$ is generated by a mixture of $N_m$ or less dynamic patterns. The $d_\mathbf{o}$-dimensional observation $\mathbf{o}_t$ is generated by a measurement model defined by $g_{\bm{\theta}_{m_t}}(\mathbf{o}_t|\mathbf{s}_t)$. The observations $\{\mathbf{o}_t\}_{t \geq 1}$ are conditionally independent given the latent process $\{\mathbf{s}_t\}_{t \geq 1}$. We follow the convention that vectors and matrices are denoted using bold fonts while scalars are denoted in normal font letters. The system diagram is shown in Fig.~\ref{fig:diagram}. 

Our goal is to jointly learn the parameter set $\bm{\theta} = \cup_{j=1}^{N_m} \bm{\theta}_j$ and track the posterior distributions $p(\mathbf{s}_{0:t}, m_{0:t}|\mathbf{o}_{1:t})$ of hidden states $\mathbf{s}_{0:t} \triangleq \{\mathbf{s}_0, \cdots, \mathbf{s}_t\}$ and model indices $m_{0:t} \triangleq \{m_0, \cdots, m_t\}$, given a collection of observations $\mathbf{o}_{1:t} \triangleq \{\mathbf{o}_1, \cdots, \mathbf{o}_t\}$.

\section{Related Work} 
\label{sec:related}

\subsection{Regime-switching particle filters}
\label{ssec:rspf}

The regime switching particle filter (RS-PF) was proposed in~\cite{el2021particle} for general regime switching systems (Equations~\eqref{eq:model_prior}-\eqref{eq:measurement_model}). The joint posterior can be factorised as~\cite{el2021particle}: 
\begin{align}
    p(\mathbf{s}_{0:t}, m_{0:t}|\mathbf{o}_{1:t}) 
    \propto & \,\, p(\mathbf{s}_{0:t-1}, m_{0:t-1}|\mathbf{o}_{1:t-1}) p(\mathbf{o}_t|\mathbf{s}_t, m_t) \nonumber \\
    &\times p(\mathbf{s}_t|\mathbf{s}_{t-1}, m_t) p(m_t|m_{0:t-1})\,. \label{equ:jointiteration}
\end{align}
The unnormalised importance weight of the $i$-th particle $\{\mathbf{s}_{0:t}^{(i)}, m_{0:t}^{(i)}\}$ is computed as: 
\begin{align}
    \bar{w}_{t}^{(i)} &= \frac{p(\mathbf{s}_{0:t}^{(i)}, m_{0:t}^{(i)}|\mathbf{o}_{1:t})}{q(\mathbf{s}_{0:t}^{(i)}, m_{0:t}^{(i)}|\mathbf{o}_{1:t})}\,, 
\end{align}
where the joint proposal distribution is factorised by: 
\begin{align}
    q(\mathbf{s}_{0:t}, m_{0:t}|&\mathbf{o}_{1:t}) = q(\mathbf{s}_{0:t-1}, m_{0:t-1}|\mathbf{o}_{1:t-1}) \nonumber \\
    & \times q(\mathbf{s}_t|\mathbf{s}_{t-1}, \mathbf{o}_t, m_t) q(m_t|m_{0:t-1})\,. 
\end{align}
If bootstrap particle filters are adopted for each candidate model, i.e., $q(\mathbf{s}_t|\mathbf{s}_{t-1}, m_t, \mathbf{o}_t) = p(\mathbf{s}_t|\mathbf{s}_{t-1}, m_t)$, the computation of the importance weight is simplified to: 
\begin{align}
    w_{t}^{(i)} &\propto \frac{p(m_{t}^{(i)}|m_{0:t-1}^{(i)}) p(\mathbf{o}_t|\mathbf{s}_t^{(i)}, m_t^{(i)})}{q(m_t^{(i)}|m_{0:t-1}^{(i)})}w_{t-1}^{(i)}\,. \label{equ:weightbootstrap}
\end{align}
Three methods were proposed to construct the regime index proposal distribution $q(m_t|m_{0:t-1})$~\cite{el2021particle}: (i) a discrete uniform distribution, i.e. $q(m_t=j|m_{1:t-1}) = \frac{1}{N_m}$ for $j=\{1,\cdots,N_m\}$; (ii) a bootstrap method, i.e. $q(m_t=j|m_{0:t-1}) = p(m_{t}|m_{0:t-1})$ for each $j$; (iii) a deterministic method that assigns an equal number of particles to each candidate model.    

\subsection{Differentiable particle filters}

Differentiable particle filters~\cite{karkus2018particle, jonschkowski2018differentiable,ma2020particle,kloss2021train,chen2021differentiable,corenflos2021differentiable} apply neural networks to construct dynamic and measurement models of particle filters in a data-adaptive way, i.e., the dynamic and measurement models are learned from data using machine learning models, e.g. neural networks. 
The forward propagation of a differentiable bootstrap particle filter~\cite{jonschkowski2018differentiable} at a single time step is illustrated in Algorithm~\ref{algo1}.  
\SetKwInput{Input}{Input}
\SetKwInput{Output}{Output}
\begin{algorithm}[htbp]
\small
    \caption{Forward propagation of a differentiable bootstrap particle filter at time step $t$}\label{algo1} 
    \Input{$\{\mathbf{s}_{t-1}^{(i)}, {w}_{t-1}^{(i)}, \bm{\epsilon}^{(i)}\}_{i=1}^{N_{p}}, \mathbf{e}_t, k_{\bm{\theta}}, l_{\bm{\theta}}$}
    \Output{$\{\mathbf{s}_{t}^{(i)}, \bar{w}_{t}^{(i)}\}_{i=1}^{N_{p}}$ \\\hrulefill} 
    Sample $\mathbf{s}_{t}^{(i)}$ according to the dynamic model $\{\mathbf{s}_{t}^{(i)} \sim f_{\bm{\theta}}(\mathbf{s}_{t}|\mathbf{s}_{t-1}^{(i)}) = k_{\bm{\theta}}(\mathbf{s}_{t-1}^{(i)}, \bm{\epsilon}^{(i)})\}_{i=1}^{N_{p}}$\; 
    Compute observation likelihood $\{g_{\bm{\theta}}(\mathbf{e}_t|\mathbf{s}_{t}^{(i)}) = l_{\bm{\theta}} (\mathbf{e}_t, \mathbf{s}_{t}^{(i)})\}_{i=1}^{N_{p}}$\; 
    Evaluate importance weights $\{\bar{w}_{t}^{(i)} = {w}_{t-1}^{(i)} g_{\bm{\theta}}(\mathbf{e}_t|\mathbf{s}_{t}^{(i)})\}_{i=1}^{N_{p}}$\; 
\end{algorithm}
Functions $k_{\bm{\theta}}(\cdot)$ and $l_{\bm{\theta}}(\cdot)$, which are the particle proposer and the observation likelihood estimator, respectively, are parameterised by neural networks. The dynamic model constructed with $k_{\bm{\theta}}(\cdot)$ takes an auxiliary noise vector $\bm{\epsilon}^{(i)}$ as part of its input for optimisation with the reparameterisation trick. The observation $\mathbf{o}_t$ can be encoded by a neural network $h_{\bm{\theta}}(\cdot)$ to generate a feature vector $\mathbf{e}_t$, i.e., $\mathbf{e}_t = h_{\bm{\theta}}(\mathbf{o}_t)$. Objective functions $\mathcal{L}(\bm{\theta})$ employed by differentiable particle filters can be classified mainly as supervised losses~\cite{karkus2018particle, jonschkowski2018differentiable, chen2021differentiable, chen2022conditional}, where the ground truth state information is available for training, and semi-supervised losses to leverage observations with unknown ground truth state information~\cite{wen2021end}.


\section{Regime-switching Differentiable Bootstrap Particle Filters} 
\label{sec:method}



We now introduce the proposed regime-switching differentiable bootstrap particle filter (RS-DBPF).
We show in Algorithm~\ref{algo2} how to integrate the regime-switching system into the design of a differentiable bootstrap particle filter. Its key steps are clarified as follows.

At the beginning of each time step, the regime index $m_t^{(i)}$ of the $i$-th particle is sampled from the model proposal distribution $q(m_{t}|m_{0:t-1})$. Three options of $q(\cdot)$ were described in the end of Section~\ref{ssec:rspf}.
Note that the filter is run with a constant number of particles $N_p$ although adaptive mechanisms could be readily used \cite{elvira2016adapting,elvira2021performance}. 
A forward propagation of the differentiable bootstrap particle filter (Algorithm~\ref{algo1}) is performed to sample the state component $\mathbf{s}_{t}^{(i)}$ with a neural network-based particle proposer $k_{\bm{\theta}_{m_{t}^{(i)}}}(\cdot)$~\cite{jonschkowski2018differentiable, karkus2018particle}, where $\bm{\theta}_{m_t^{(i)}}$ denotes the parameter set of the $m_t^{(i)}$-th candidate model:
\begin{align}
    \mathbf{s}_{t}^{(i)} \sim f_{\bm{\theta}_{m_t^{(i)}}}(\mathbf{s}^{(i)}_{t}|\mathbf{s}_{t-1}^{(i)}) = k_{\bm{\theta}_{m_t^{(i)}}}(\mathbf{s}_{t-1}^{(i)}, \bm{\epsilon}^{(i)}) \label{equ:sampling}\,.
\end{align}
The auxiliary noise term $\bm{\epsilon}^{(i)} \sim \mathcal{N}(\mathbf{0}_{d_\mathbf{s}}, \mathbf{I}_{d_\mathbf{s}})$ where $\mathbf{0}_{d_\mathbf{s}}$ denotes a $d_\mathbf{s}$-dimensional zero vector and $\mathbf{I}_{d_\mathbf{s}}$ is a $d_{\mathbf{s}}\times d_{\mathbf{s}}$ identity matrix. The likelihood of the $i$-th particle is computed as~\cite{jonschkowski2018differentiable, karkus2018particle}: 
\begin{align}
    p(\mathbf{o}_t|\mathbf{s}_t^{(i)}, m_t^{(i)}) &= g_{\bm{\theta}_{m_t^{(i)}}}(\mathbf{o}_t|\mathbf{s}_{t}^{(i)}) = l_{\bm{\theta}_{m_t^{(i)}}}(\mathbf{o}_t,\mathbf{s}_t^{(i)}) \label{equ:measurement}\,.
\end{align}

Assuming knowledge of the model switching distribution $p(m_{t}^{(i)}|m_{1:t-1}^{(i)})$, the unnormalised importance weight $\bar{w}_{t}^{(i)}$ is updated following Equation~\eqref{equ:weightbootstrap}:
\begin{align}
\bar{w}_{t}^{(i)} = {w}_{t-1}^{(i)}\frac{p(m_{t}^{(i)}|m_{1:t-1}^{(i)}) g_{\bm{\theta}_{m_t^{(i)}}}(\mathbf{o}_t|\mathbf{s}_{t}^{(i)})}{q(m_t^{(i)}|m_{1:t-1}^{(i)})}\,.
\end{align}

\begin{algorithm}[htbp]
\small
    \caption{Regime switching differentiable bootstrap particle filters (RS-DBPFs) framework}\label{algo2}
    \Input{$\eta$, $\pi(m_0)$, $\mu(\mathbf{s}_0)$, $N_p$, $T$, $\mathbf{o}_{1:T}$, $k_{\bm{\theta}}$, $l_{\bm{\theta}}$, $\text{ESS}_{\text{thres}}$, $\mathcal{L}$}
    Initialise parameter sets $\bm{\theta}_j \subseteq \bm{\theta}$ for $j=\{1,\cdots,N_m\}$ and set learning rate $\eta$\;  
    \While{$\bm{\theta}$ has not converged}{
        Draw regime index $\{{m}_{0}^{(i)}\}_{i=1}^{N_p} \sim \pi(m_{0})$\; 
        Draw samples $\{\mathbf{s}_{0}^{(i)}\}_{i=1}^{N_p} \sim \mu (\mathbf{s}_{0})$\;
        Set importance weights $\{{w}_{0}^{(i)}\}_{i=1}^{N_p} = \frac{1}{N_{p}}$\; 
        \For{$t = 1, 2,\cdots, T$}{
            Draw regime index from the model proposal distribution $\{m_t^{(i)} \sim q(m_t|{m}_{1:t-1}^{(i)})\}_{i=1}^{N_p}$\; 
            Sample $\mathbf{s}_{t}^{(i)}$ as in  Eq.~\eqref{equ:sampling}\; 
            Estimate observation likelihood $g_{\bm{\theta}_{m_t^{(i)}}}(\mathbf{o}_t|\mathbf{s}_{t}^{(i)})$ according to Equation~\eqref{equ:measurement}\; 
            Calculate importance weights $\{\bar{w}_{t}^{(i)} = {w}_{t-1}^{(i)}\frac{p(m_{t}^{(i)}|m_{1:t-1}^{(i)}) g_{\bm{\theta}_{m_t^{(i)}}}(\mathbf{o}_t|\mathbf{s}_{t}^{(i)})}{q(m_t^{(i)}|m_{1:t-1}^{(i)})}\}_{i=1}^{N_p}$\; 
            Normalise weights $\{{w}_{t}^{(i)} = \frac{\bar{w}_{t}^{(i)}}{\sum_{k=1}^{N_p} \bar{w}_{t}^{(k)}}\}_{i=1}^{N_p}$\; 
            Compute the effective sample size $\text{ESS}_{t} = \frac{1}{\sum_{i=1}^{N_p} ({w}_{t}^{(i)})^2}$\; 
            \If{$\textup{ESS}_t < \textup{ESS}_{\textup{thres}}$}{
                Resample and update $\{m_{1:t}^{(i)}, \mathbf{s}_{1:t}^{(i)}\}_{i=1}^{N_{p}}$ according to importance weights ${w}_{t}^{(i)}$\;
                Update $\{{w}_{t}^{(i)} = \frac{1}{N_p}\}_{i=1}^{N_{p}}$\;
            }
            Calculate the estimate $\hat{\mathbf{s}}_{t} = \sum_{i=1}^{N_{p}} {w}_{t}^{(i)} \mathbf{s}_{t}^{(i)}$\; 
        }
        Calculate the total loss $\mathcal{L}(\bm{\theta})$\; 
        Update parameters by gradient descent $\bm{\theta}_j = \bm{\theta}_j - \eta \triangledown_{\bm{\theta}_j} \mathcal{L}$ for $j=\{1,\cdots,N_m\}$\; 
    }
\end{algorithm} 

Finally, a resampling step is performed if the effective sample size (ESS) is smaller than a threshold \cite{elvira2022rethinking}.

\section{Simulations and Results}
\label{sec:experiment}

We adopt the synthetic data experiment explored in~\cite{el2021particle}.
It includes a mixture of eight candidate dynamic and measurement models with small variances for each candidate model. This leads to multi-modal posterior distributions that are challenging for filtering algorithms to explore all modes\footnote{Code to reproduce experiment results is available at \url{https://github.com/WickhamLi/RS-DBPF}}.

\subsection{Experiment setting}

The $j$-th candidate model ($j\in\{1,\cdots,8\}$) is as follows: 
\begin{equation}
\mathcal{M}_j: 
\left\{
    \begin{array}{lr}
        s_{t} = a_j s_{t-1} + b_j + u_t \\ 
        o_{t} = c_j \sqrt{\left\lvert s_{t}\right\rvert} + d_j +v_t 
    \end{array}
    \,. 
\right.
\end{equation}
The number of time steps for one trajectory is $T=50$. The initial state $s_0\in\mathbb{R}$ is sampled from a continuous uniform distribution $\mathcal{U}{[-0.5,0.5]}$ whereas the index of the initial sub-model $m_0$ is sampled from a discrete uniform distribution $\mathcal{U}\{1,8\}$. Coefficients $[a_1,\cdots,a_8] = [-0.1, -0.3, -0.5, -0.9, 0.1, 0.3, 0.5, 0.9]$, $[b_1,\cdots,b_8] = [0, -2, 2, -4, 0, 2, -2, 4]$, $[c_1,\cdots,c_8] = [a_1,\cdots,a_8]$, and $[d_1,\cdots,d_8] = [b_1,\cdots,b_8]$. The noise terms $u_{t} \sim \mathcal{N}(0, 0.1)$ and $v_{t} \sim \mathcal{N}(0, 0.1)$.  The overall dataset includes $2000$ trajectories ($1000$ for training, $500$ for validation, $500$ for testing). 

The regime switching dynamic follows either a Markovian dynamic or a P{\'o}lya urn dynamic. 
In the Markovian switching system, $p(m_{t}|m_{0:t-1}) = p(m_{t}|m_{t-1})$. Following the example in~\cite{el2021particle}, we set the transition probability matrix $\mathbf{P}$ as: 
\begin{align}
    \mathbf{P} = 
    \begin{bmatrix}
        0.80 & 0.15 & \rho & \cdots & \rho \\
        \rho & 0.80 & 0.15 & \cdots & \rho \\
        \vdots &  & \ddots & & \vdots \\
        \rho & \cdots &  \rho & 0.80 & 0.15 \\
        0.15 & \rho & \cdots & \rho & 0.80
    \end{bmatrix}\,,
\end{align}
where $\rho = \frac{1}{120}$. 
$\mathbf{P}_{j, k} \triangleq p(m_{t}=k|m_{t-1}=j)$. 


The P{\'o}lya switching model is a more general dynamic process to describe long-term time dependencies between candidate models:
\begin{align}
    p(m_{t}=k|m_{0:t-1}) = \frac{\sum_{\tau=0}^{t-1} \mathds{1}_{k, \tau} + \beta_{k}}{\sum_{j=1}^{N_m} (\sum_{\tau=0}^{t-1}\mathds{1}_{j, \tau} + \beta_{j})} \,, 
\end{align}
where $k = \{1,\cdots, N_m\}$ denotes the regime index, $\mathds{1}_{k, \tau} = \mathds{1} (m_{\tau} = k)$ is an indicator function on whether the system is switched to the $k$-th model at time step $\tau$. $\beta_{k}$ is set to $1$.

\subsection{Parameter values for the filtering algorithms}

We compare the proposed regime-switching differentiable bootstrap particle filter (RS-DBPF) with a multi-model particle filter (MM-PF)~\cite{liu2011instantaneous}, a differentiable bootstrap particle filter (DBPF)~\cite{jonschkowski2018differentiable}, and a regime-switching particle filter (RS-PF)~\cite{el2021particle}. $200$ particles are employed for training and validation to reduce computational costs. $2000$ particles are used to perform filtering with testing trajectories for smooth estimated trajectories. Particles are initialised from a uniform distribution $\mathcal{U}[-0.5,0.5]$. The RS-DBPF and the RS-PF adopt a uniform distribution for the regime index proposal $q(m_t|m_{1:t-1})$ to evaluate the robustness of the filters when the regime index proposal deviates from the regime switching dynamic.

For the DBPF and the RS-DBPF, the particle proposer $k_{\bm{\theta}}$ adopts a $2$-layer neural network with $8$ neurons in the hidden layer. A Gaussian kernel with a learnable kernel parameter is used to generate conditional likelihoods $l_{\bm{\theta}}$, by comparing the observation with an embedding generated from the state through a $2$-layer neural network with $8$ neurons in the hidden layer with a tanh-activation function.  A supervised loss based on mean squared errors between the ground truth and predicted states is adopted. Stochastic gradient descent with a momentum factor of $0.9$ is used as optimiser. We choose learning rates $\eta\in\{0.01, 0.02, 0.05, 0.1\}$ with a step-wise decay that halves the learning rate every $10$ epochs for the DBPF and the RS-DBPF based on best validation performance. The epoch number is set to $60$. The mini-batch size is set to $100$.



\begin{table}[htbp]
\caption{Average, best, and worst RMSEs with a Markovian dynamic.}
\begin{center}
\begin{tabular}{|c|c|c|c|}
\hline
 & {\textbf{Average}} & {\textbf{Best}} & {\textbf{Worst}} \\
\hline
MM-PF (baseline) & 1.9016 & 0.5601 & 9.4422  \\
\hline
DBPF (baseline) & 1.5176 & 0.5085 & 9.5790 \\
\hline
RS-DBPF (proposed) & 0.8325 & 0.3779 & 8.6401 \\
\hline
\hline
RS-PF (oracle) & 0.4627 & 0.2570 & 2.2972 \\
\hline
\end{tabular}
\label{tab:Mark}
\end{center}
\end{table}
\begin{table}[tbhp]
\caption{Average, best, and worst RMSEs with a P{\'o}lya dynamic.}
\begin{center}
\begin{tabular}{|c|c|c|c|}
\hline
\footnotesize
 & {\textbf{Average}} & {\textbf{Best}} & {\textbf{Worst}}\\
\hline
MM-PF (baseline) & 2.1334 & 0.6409 & 5.3060\\
\hline
DBPF (baseline) & 1.6144 & 0.4754 & 5.4350 \\
\hline
RS-DBPF (proposed) & 0.8394 & 0.3817 & 2.5627\\
\hline
\hline
RS-PF (oracle) & 0.6399 & 0.3171 & 2.0383\\ 
\hline
\end{tabular}
\label{tab:Poly}
\end{center}
\end{table}

\subsection{Tracking performance}

\begin{figure}[htbp]
\centerline{\includegraphics[width=9cm]{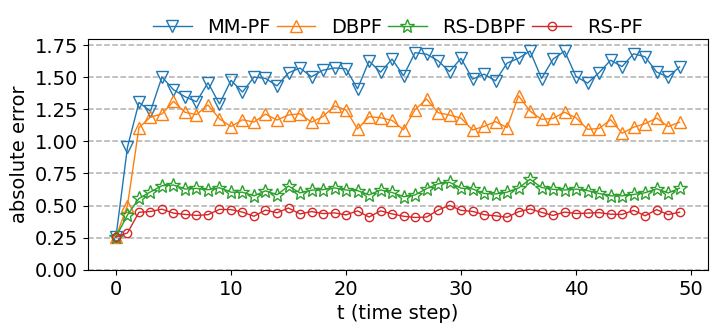}}
\caption{Mean absolute errors at each time step (P{\'o}lya urn dynamics).}
\label{fig:MSEPoly}
\end{figure}

We compute root mean squared errors (RMSEs) of predicted states for each test trajectory and report error statistics among $500$ test trajectories in TABLEs~\ref{tab:Mark} and~\ref{tab:Poly}. The proposed RS-DBPF leads to significantly smaller average RMSEs compared with the baselines including the DBPF and the MM-PF. Fig.~\ref{fig:MSEPoly} plots the absolute errors along each time step averaged over all the test trajectories for the P{\'o}lya urn switching model. Note that both the RS-DBPF and the DBPF assume no knowledge of the candidate models, while the MM-PF and the RS-PF have access to the ground truth candidate models. The RS-PF can further utilise the regime switching dynamic so it is served as the oracle model to generate optimal filtering performance as a benchmark. The poor performance of the MM-PF is due to its algorithmic assumption that there is no regime switching.

\section{Conclusion}
\label{sec:conclusion}

In this paper, we address filtering tasks where a mixture of unknown candidate dynamic and measurement models exist. The proposed RS-DBPF can flexibly switch between candidate models, i.e. regimes, while simultaneously learn the candidate models without prior knowledge of their functional forms. Numerical simulations show that the RS-DBPF outperforms both a vanilla DBPF and a MM-PF in two simulation setups with different regime switching dynamics. Future work includes the estimation of the regime-switching dynamic, the incorporation of more expressive neural networks to construct candidate models, and more extensive experimental evaluation with high-dimensional numerical and real-world experiments.


\clearpage
\vfill\pagebreak

\bibliographystyle{IEEEtran}
\bibliography{refs}

\end{document}